\documentclass[12pt]{iopart}

\usepackage{iopams} 
\usepackage{bm,graphics,graphicx,epsfig,color}

\begin{document}

\title[Carter-like constants of motion]{Carter-like constants of motion in the Newtonian and relativistic two-center problems}

\author{Saeed Mirshekari$^{1}$ and Clifford M. Will$^{1,2}$}

\ead{smirshekari@wustl.edu, cmw@wuphys.wustl.edu}
\address{
$^1$ McDonnell Center for the Space Sciences, Department of Physics,
Washington University, St. Louis MO 63130 USA
 \\
$^2$ GReCO, Institut d'Astrophysique de Paris, UMR 7095-CNRS,
Universit\'e Pierre et Marie Curie, 98$^{bis}$ Bd. Arago, 75014 Paris, France
}

\begin{abstract}
In Newtonian gravity, a stationary axisymmetric system admits a third, Carter-like constant of motion if its mass multipole moments are related to each other in exactly the same manner as for the Kerr black-hole spacetime.  The Newtonian source with this property consists of two point masses at rest a fixed distance apart.  The integrability of motion about this source was first studied in the 1760s by Euler.  We show that the general relativistic analogue of the Euler problem, the Bach-Weyl solution, does not admit a Carter-like constant of motion, first, by showing that it does not possess a non-trivial Killing tensor, and secondly, by showing that the existence of a Carter-like constant for the two-center problem fails at the first post-Newtonian order.

\end{abstract}

\maketitle

\section{Introduction and summary}

Every physics undergraduate knows that solving dynamical problems in classical mechanics is simplified if one can find constants of the motion.  Given enough suitable constants of motion, the problem becomes completely integrable, solvable in terms of integrals.  The simplest example is a static, spherically symmetric potential.  In Newtonian gravitation, the conservation of energy, total angular momentum and one of its components guarantees that the motion for this problem is completely integrable.  Likewise, in general relativity, given the same symmetries, the conservation of the rest mass (or the unit norm of the four velocity) in addition to the other three constants of motion leads to an integrable system.  But for problems with less symmetry, completely integrable systems are rare.

Systems that are stationary and axisymmetric possess conserved energy and angular momentum about the symmetry axis (and conserved rest mass in the relativistic case).  But under what conditions do they possess an additional constant of the motion, sufficient to make them integrable? 
Recently, one of us~\cite{will09} pointed out that a stationary axisymmetric system in Newtonian gravitational theory could possess such a third constant of the motion, analogous to the Carter constant of the Kerr geometry, if the multipole moments $Q_{\ell}$ of the source obey the {\em same} constraints as those satisfied by the mass moments of the Kerr spacetime, namely
\begin{equation}
Q_{2\ell} = m a^{2\ell} \,, \quad  Q_{2\ell+1} = 0\,,
\label{eq1}
\end{equation}
where $\ell = 0, \, 1,\, 2,\, \dots$.
Unlike the Kerr case, where $a$ is the angular momentum per unit mass, here
$a \equiv (Q_2/m)^{1/2}$, where $Q_2$ is the quadrupole moment and $m$ is the total mass of the system.  Nevertheless the similarity of Eq.\ (\ref{eq1}) to the Kerr sequence of mass moments is striking.   

In the prolate case, $Q_2 >0$, the Newtonian source that has this property consists of two bodies each of mass $m/2$ held at fixed locations $\pm a$  on the $z$-axis.  Subsequent to the publication of Ref.\ \cite{will09}, its author learned that this problem was already studied in the 18th century by none other than Leonhard Euler, and is now known to classical dynamicists as the ``Euler problem'', a non-spherically symmetric, yet completely integrable system (for a thorough review of the history and mathematics of the Euler problem and related integrable systems, see \cite{mathuna}).   Other authors have similarly rediscovered the unique properties of the Euler problem in other contexts, including the quantum mechanics of the singly ionized hydrogen molecule~\cite{coulson} (the underlying electrodynamics is mathematically equivalent to the gravitational Euler problem) and stellar dynamics~\cite{lyndenbell}. 

In the oblate case $Q_2 <0$, $a$ is imaginary, and the source consists formally of two equal point masses at $\pm {\rm i} a$ on the imaginary $z$-axis, although the gravitational potential is still perfectly real.

It turns out that the solution of Ref.\ \cite{will09} can be generalized to the case where the dipole moment $Q_1$ is non-zero; here the sources in the prolate case are two {\em unequal} masses held at $\pm a$ on the $z$-axis, with $m_+ + m_- = m$.   In the oblate case, the sources are two complex conjugate masses on the imaginary $z$-axis.   Some of these generalizations had previously been studied by celestial mechanician John P. Vinti \cite{vinti}, who noticed that if the parameter $a$ of the Euler problem was adjusted to match the Earth's $Q_2$, then the value of $Q_4$ implied by Eq.\ (\ref{eq1}) closely approximated that of the Earth (though higher moments did not), so that the integral solutions of the Euler problem gave reasonable approximations for orbits of satellites around the Earth \cite{mathuna}.  

These results in Newtonian gravity raise the obvious question: is there an analogous third integral of the motion for the general relativistic version of the Euler problem, namely the Bach-Weyl solution \cite{weyl,bachweyl}?  This is an exact, static, axisymmetric, vacuum solution of Einstein's equations, corresponding to two point masses held a fixed distance apart on the $z$-axis.  It is asymptotically flat, and is regular everywhere except along a line joining the two masses; on this line there is a pressure singularity representing a ``strut'' needed to hold the masses apart.

In this paper, we show that in fact the Bach-Weyl solution does {\em not} admit a Carter-like constant analogous to that of the Newtonian Euler problem.  We do this in two ways.  First we show that the Bach-Weyl spacetime does not admit a non-trivial second-rank Killing tensor $\xi_{\alpha\beta}$.  Had it done so, then by virtue of the Killing equations $\xi_{(\alpha\beta; \gamma)} =0$ and the geodesic equation $p^\beta {p^\alpha}_{;\beta}=0$, there would have existed a conserved quantity given by $C = \xi_{\alpha\beta} p^\alpha p^\beta$.   It is important to recall that the metric itself and any product of Killing vectors are also Killing tensors, but they are trivial in that they do not generate new constants of the motion.  

Second, we show that, in the post-Newtonian limit of the Bach-Weyl geometry, a Carter-like constant analogous to that which exists at Newtonian order cannot be found.   Whereas in the Newtonian Euler problem, the Carter-like constant is given by
\begin{equation}
C = h^2 + a^2 v_z^2 - 2azU^* \,.
\label{sec1:carter}
\end{equation}
where ${\bm h} = {\bm x} \times {\bm v}$ is the angular momentum per unit mass, and 
\begin{equation}
U^* \equiv \frac{Gm_+}{r_+}-\frac{Gm_-}{r_-}  \,, 
\end{equation}
where $r_{\pm} = |{\bm x} \mp a {\bm e}_z |$ and
$m_{\pm}$ are the two masses (in general they do not need to be equal), in the post-Newtonian Euler problem, we show that the 
closest one can get to a conserved quantity is the expression
\begin{equation}
 \frac{1}{2} \frac{d}{dt} \left [ (h^2 + a^2 v_z^2)\left (1+  \frac{8U}{c^2} \right)
-2azU^* \left (1+ \frac{v^2}{c^2} +\frac{4U}{c^2} \right) \right ] = -\frac{6}{c^2} azU^* \frac{dU}{dt} \,,
\end{equation}
where $U = Gm_+/r_+ +Gm_-/r_-$. 
The term on the right-hand side cannot be expressed as a total time derivative of any function of $\bm x$ or $\bm v$, and so a Carter-like constant of the chosen form does not exist at post-Newtonian order.  

The remainder of this paper gives the details underlying these conclusions.  In Sec.\ II we discuss the Bach-Weyl solution and its symmetries.  In Sec.\ III we derive the general solution of the Killing tensor equations, and show that the result is a linear combination of trivial Killing tensors.  Section IV takes the post-Newtonian limit of the Bach-Weyl solution and shows by construction that a post-Newtonian generalization of the Euler-problem's Carter-like constant does not exist.  Concluding remarks are made in Sec.\ V.  In an Appendix, we give a more detailed argument supporting a key step in the analysis.

\section{The Bach-Weyl solution}

The Bach-Weyl metric \cite{weyl,bachweyl,stephani} is given in cylindrical $\rho -z$ coordinates  by
\begin{equation}\label{metricds}
ds^2= -e^{2\lambda}dt^2+ e^{-2\lambda}[e^{2\nu}(d\rho^2+dz^2)+\rho^2d\phi^2] \,,
\end{equation}
where
\begin{eqnarray}\label{lambda}
\lambda(\rho,z)&=&-\frac{m_+}{r_+}-\frac{m_-}{r_-}  \,,
\\
\label{nu}
\nu(\rho,z)&=&-\frac{\rho^2}{2} \left (\frac{m_+^2}{r_+^4}+\frac{m_-^2}{r_-^4} \right )+\frac{m_+m_-}{2a^2} \left (\frac{\rho^2+z^2-a^2}{r_+ r_-} \right ) \,,
\end{eqnarray}
where $r_\pm^2 =\rho^2+(z\mp a)^2$.  Here and henceforth, we use units in which $G=c=1$.  The total Kepler mass of the system is given by $m_+ + m_-$. 
The physical system can be viewed as consisting of two point masses at $z=-a$ and $z=+a$; the metric is regular everywhere except at the two points, but the Einstein tensor (and thus the stress energy tensor) diverges along a line joining the two points,  representing the ``strut'' required to hold the two masses a fixed distance apart.

The metric admits the timelike and azimuthal Killing vectors, 
${\bm \xi}^{(t)} = \partial / \partial t$ and ${\bm \xi}^{(\phi)} = \partial / \partial \phi$.  
If it also admits a symmetric, second-rank Killing tensor $\xi_{\alpha\beta}$, then $\xi_{\alpha\beta}$ satisfies the equations \cite{stephani}
\begin{equation}
K_{\alpha\beta\gamma} \equiv \xi_{\gamma\beta ; \alpha}+\xi_{\alpha\gamma ; \beta}+\xi_{\alpha\beta ; \gamma}=0 \,,
\label{kteq}
\end{equation}
where Greek indices run over the values $0 \dots 3$, semicolons in subscripts denote covariant derivatives, and
the notation $K_{\alpha\beta\gamma}$  labels the partial differential equations.   The metric $g_{\alpha\beta}$ and any symmetrized product ${\xi^{(A)}}_{(\alpha} {\xi^{(B)}}_{\beta)}$ of Killing vectors automatically satisfy the Killing tensor equations, and thus our goal is to find non-trivial solutions of Eq.\ (\ref{kteq}).  Only a non-trivial Killing tensor will generate a new constant $\xi_{\alpha\beta} p^\alpha p^\beta$ of the motion.

\section{Searching for a non-trivial Killing tensor}

We define a template for the symmetric Killing tensor as 
\begin{equation}
\xi_{\alpha \beta} = \left(
\begin{array}{cccc}
   A&F&G&Q\\
   F&D&E&H\\
   G&E&C&J\\
   Q&H&J&B
\end{array}
\right) \,,
\label{ktensor}
\end{equation}
where the 10 functions $A, \dots , Q$ are {\em a priori} functions of $t, \,\rho,\, z$ and  $\phi$.  Notice that, notwithstanding the fact that the metric is stationary and axisymmetric, it is incorrect to assume {\em a priori} that the Killing tensor is independent of $t$ and $\phi$; in spherical symmetry for example, the trivial Killing tensors constructed from products of the two non-axial rotational Killing vectors depend explicitly on $\phi$.  Using Eqs.\ (\ref{metricds}) and  (\ref{ktensor}) in Eq.\ (\ref{kteq}), we obtain a system of 20 coupled partial differential equations.  Following a method suggested by Garfinkle and Glass \cite{garfinkle} we divide the equations into three groups corresponding to the $(\rho,\,z)$ block  (the functions $D$, $C$ and $E$), the $(\rho,\,z,\,\phi)$ block ($H$, $J$ and $B$) and the remaining $(t,\, \rho,\,z,\,\phi)$ block ($A$, $F$, $G$ and $Q$).   We define $\alpha = \lambda - \nu$.

\noindent
(a) $(\rho, \, z)$ block:
\begin{eqnarray}
K_{\rho z z}:\, 0&=&   \frac{\partial C}{\partial \rho}
+2\frac{\partial E}{\partial z}
+2(2C-D) \frac{\partial \alpha}{\partial \rho}+ 6E  \frac{\partial \alpha}{\partial z} \,,
\label{Krzz}\\
K_{z \rho \rho}:\, 0&=& \frac{\partial D}{\partial z}
+2\frac{\partial E}{\partial\rho}
+ 2(2D-C)\frac{\partial \alpha}{\partial z} + 6E \frac{\partial \alpha}{\partial \rho} \,,
\label{Kzrr}\\
K_{\rho \rho \rho}:\, 0&=&  \frac{\partial D}{\partial \rho}
+ 2D \frac{\partial \alpha}{\partial \rho}-2E \frac{\partial \alpha}{\partial z} \,,
\label{Krrr}\\
K_{z z z}:\, 0&=&  \frac{\partial C}{\partial z}+2C \frac{\partial \alpha}{\partial z}-2E  \frac{\partial \alpha}{\partial \rho} \,.
\label{Kzzz}
\end{eqnarray}

\noindent
(b) $(\rho, \, z, \, \phi)$ block:
\begin{eqnarray}
K_{\phi\rho\rho}:\, 0&=&  \frac{\partial H}{\partial\rho}+H\left (2\frac{\partial\lambda}{\partial\rho}+\frac{\partial\alpha}{\partial\rho} \right )-J\frac{\partial\alpha}{\partial z}-2\frac{H}{\rho}+\frac{1}{2}\frac{\partial D}{\partial \phi} \,,
\label{Kprr}\\
K_{\phi z z}:\, 0&=&  \frac{\partial J}{\partial z}-H\frac{\partial\alpha}{\partial\rho}+J \left (2\frac{\partial\lambda}{\partial z}+\frac{\partial\alpha}{\partial z} \right )+\frac{1}{2}\frac{\partial C}{\partial \phi} \,,
\label{Kpzz}\\
K_{\phi z\rho}:\, 0&=&  
\frac{\partial H}{\partial z}+\frac{\partial J}{\partial\rho} +
2H \left (\frac{\partial\lambda}{\partial z}+\frac{\partial\alpha}{\partial z} \right  )
\nonumber \\
&&+2J \left (\frac{\partial\lambda}{\partial\rho}+\frac{\partial\alpha}{\partial\rho} -\frac{1}{\rho} \right )+\frac{\partial E}{\partial \phi} \,,
\label{Kpzr}\\
K_{\rho \phi \phi}:\, 0&=&  e^{2\nu} \left ( \frac{\partial B}{\partial \rho} +4 B \frac{\partial \lambda}{\partial \rho}-4\frac{B}{\rho}+2\frac{\partial H}{\partial \phi} \right )
\nonumber \\
&&-2D\rho^2 \left ( \frac{\partial \lambda}{\partial \rho}-\frac{1}{\rho} \right )-2E\rho^2\frac{\partial\lambda}{\partial z} \,,
\label{Krpp}\\
K_{z \phi \phi}:\, 0&=&  e^{2\nu} \left (\frac{\partial B}{\partial z}+4B\frac{\partial\lambda}{\partial z}+2\frac{\partial J}{\partial\phi} \right )-2C\rho^2\frac{\partial\lambda}{\partial z}
\nonumber \\
&&
-2E\rho^2
\left (\frac{\partial\lambda}{\partial\rho}-\frac{1}{\rho} \right ) \,,
\label{Kzpp}\\
K_{\phi\phi\phi}:\, 0&=&  H \left (\frac{\partial\lambda}{\partial\rho}-\frac{1}{\rho} \right )+J\frac{\partial\lambda}{\partial z}-\frac{1}{2}\rho^{-2}\frac{\partial B}{\partial\phi}e^{2\nu} \,.
\label{Kppp}
\end{eqnarray}

\noindent
(c) $(\rho, \, z, \, \phi, \, t)$ block:
\begin{eqnarray}
K_{\rho\rho t}:\, 0&=& \frac{\partial F}{\partial\rho}
-F \left (\frac{\partial\lambda}{\partial\rho}+\frac{\partial\nu}{\partial\rho} \right )
-G \left (\frac{\partial\lambda}{\partial z}-\frac{\partial\nu}{\partial z} \right ) 
+ \frac{1}{2} \frac{\partial D}{\partial t} \,,
\label{Krrt} \\
K_{z z t} :\, 0&=& \frac{\partial G}{\partial z}
-G \left (\frac{\partial\lambda}{\partial z}+\frac{\partial\nu}{\partial z} \right )
-F \left (\frac{\partial\lambda}{\partial\rho}-\frac{\partial\nu}{\partial\rho} \right ) 
+\frac{1}{2} \frac{\partial C}{\partial t}\,,
\label{Kzzt} \\
K_{\rho z t}:\, 0&=&  \frac{\partial G}{\partial\rho}+\frac{\partial F}{\partial z}-2F\frac{\partial\nu}{\partial z}-2G\frac{\partial\nu}{\partial\rho} + \frac{\partial E}{\partial t} \,,
\label{Krzt} \\
K_{\rho t t}:\, 0&=& \frac{\partial A}{\partial\rho}-4A\frac{\partial \lambda}{\partial \rho}-2e^{4\lambda-2\nu } \left ( D\frac{\partial \lambda}{ \partial \rho}+E\frac{\partial\lambda}{\partial z} \right )
+2\frac{\partial F}{\partial t} \,,
\label{Krtt}\\
K_{z t t}:\, 0&=& \frac{\partial A}{\partial z}-4A\frac{\partial \lambda}{\partial z}-2 e^{4\lambda-2\nu} \left (C\frac{\partial \lambda}{ \partial z}+E\frac{\partial\lambda}{\partial\rho} \right )
+2\frac{\partial G}{\partial t}\,,
\label{Kztt} \\
K_{ttt}:\, 0&=&  - \frac{1}{2} \frac{\partial A}{\partial t} e^{2\nu-4\lambda}+F\frac{\partial\lambda}{\partial\rho}+G\frac{\partial\lambda}{\partial z} \,,
\label{Kttt}\\
K_{\phi\rho t}:\, 0&=&  \frac{\partial H}{\partial t}+\frac{\partial Q}{\partial\rho}-2\frac{Q}{\rho}+\frac{\partial F}{\partial \phi} \,,
\label{Kprt}\\
K_{\phi z t}:\, 0&=&  \frac{\partial J}{\partial t}+\frac{\partial Q}{\partial z}+\frac{\partial G}{\partial \phi} \,,
\label{Kpzt} \\
K_{\phi t t}:\, 0&=&  2\frac{\partial Q}{\partial t}e^{2\nu}-2e^{4\lambda} \left (H\frac{\partial\lambda}{\partial\rho}+J\frac{\partial\lambda}{\partial z} \right )+\frac{\partial A}{\partial \phi}e^{2\nu} \,,
\label{Kptt} \\
K_{t\phi\phi}:\, 0&=& F \left (\frac{\partial\lambda}{\partial\rho} -\frac{1}{\rho} \right ) +G\frac{\partial\lambda}{\partial z}-\rho^{-2}\frac{\partial Q}{\partial \phi}e^{2\nu}  -\frac{1}{2}\rho^{-2}\frac{\partial B}{\partial t}e^{2\nu}\,.
\label{Ktpp}
\end{eqnarray}

First, we focus on the $\rho,\, z$ block, Eqs.\ (\ref{Krzz}) - (\ref{Kzzz}).  After changing variables to 
\begin{equation}\label{change1}
C \equiv W V + Y V^2\,, \quad  D \equiv W V - Y V^2\,, \quad E \equiv X V^2 \,,
\end{equation}
where $V\equiv e^{-2\alpha}$, and $W$, $X$ and $Y$ are functions of $t, \,\rho,\, z$ and  $\phi$, we obtain%
\begin{eqnarray}
\label{rhozz}
K_{\rho zz}:\, 0&=&   W_{,\rho}+Y_{,\rho}V - YV_{,\rho}+ 2X_{,z}V+XV_{,z} \,,
\\
\label{zrhorho}
K_{z \rho \rho}:\, 0&=&  W_{,z}-Y_{,z}V + YV_{,z}+ 2X_{,\rho}V+XV_{,\rho} \,,
\\
\label{rhorhorho}
K_{\rho \rho \rho}:\, 0&=&    W_{,\rho}-Y_{,\rho}V - YV_{,\rho}+XV_{,z} \,,
\\ 
\label{zzz}
K_{z z z}:\, 0&=&  W_{,z}+Y_{,z}V + YV_{,z}+XV_{,\rho} \,,
\end{eqnarray}
where commas in subscripts denote partial derivatives.
Subtracting Eq.\ (\ref{rhorhorho}) from Eq.\ (\ref{rhozz}) and Eq.\ (\ref{zrhorho}) from Eq.\ (\ref{zzz}) yields
\begin{eqnarray}
Y_{,\rho}+X_{,z}=0 \,, \quad
Y_{,z}-X_{,\rho}=0 \,,
\label{YXequations}
\end{eqnarray}
whose general solution is
\begin{eqnarray}
Y+iX=\mathcal{F}(\xi,\phi,t) \,,
\end{eqnarray}
where $\mathcal{F}$ is an arbitrary analytic function of the complex variable
$\xi=z+i\rho =re^{i\theta}$, and an arbitrary function of the variables $\phi$ and $t$. Notice that $\mathcal{F}$ is proportional to the variable $t$ defined by Brink~\cite{brink}.   

Taking the first derivative of Eqs.\ (\ref{rhozz}) and (\ref{zrhorho}) with respect to $z$ and $\rho$ respectively, subtracting, and using the fact that $\partial/\partial \rho = i (\partial/\partial \xi - \partial/\partial {\bar \xi} )$ and $\partial/\partial z = \partial/\partial \xi + \partial/\partial {\bar \xi}$, we obtain
\begin{equation}
V(\xi,{\bar \xi})\mathcal{F}_{,\xi\xi}(\xi,\phi,t)
+3V_{,\xi}(\xi,{\bar \xi})\mathcal{F}_{,\xi}(\xi,\phi,t) + 2V_{,\xi\xi}(\xi,{\bar \xi})\mathcal{F}(\xi,\phi,t)
= {\rm c.c.}  \,,
\label{Fequation}
\end{equation}
where c.c.\ denotes the complex conjugate of the left-hand-side.
This equation can also be written in the form
\begin{equation}
\left ( \frac{(\mathcal{F} V^2)_{,\xi} }{V} \right )_{,\xi} = {\rm c.c.} \,.
\label{Fequation2}
\end{equation}
This is equivalent to Brink's Eq.\ (16).

Because $V= e^{-2\alpha(\rho, z)}$ is a complicated, non-analytic function of $\rho$ and $z$ (or of $\xi$ and $\bar \xi$), we can conclude that Eq.\ (\ref{Fequation}) or (\ref{Fequation2}) can be satisfied for arbitrary $\rho$ and $z$ if and only if
\begin{equation}
\mathcal{F}(\xi,\phi,t)=0 \,,
\label{calF0}
\end{equation}
which immediately gives
\begin{eqnarray}\label{heart}
X=Y= 0 \,.
\end{eqnarray}
A more detailed argument supporting this claim is presented in the Appendix.
From Eqs.\ (\ref{rhozz}) and (\ref{zrhorho}) we then conclude that
$W$  is purely a function of $\phi$ and $t$ i.e.
$W=C_1(\phi,t)$, and thus that
\begin{equation}
\label{CDE}
C=D=C_1(\phi,t)e^{-2\alpha} \,, \qquad E=0 \,.
\end{equation}

Armed with the solution for $C$, $D$, and $E$, we now consider Eqs.\ (\ref{Kprr}) - (\ref{Kppp}) involving the functions $H$, $J$ and $B$.   
After changing variables to
\begin{eqnarray}
\label{change2}
J \equiv j\rho^2 e^{-2(2\lambda-\nu)}, & H \equiv h\rho^2e^{-2(2\lambda-\nu)}, & B \equiv C_1(\phi,t)\rho^2e^{-2\lambda}+C_3\rho^4e^{-4\lambda},
\end{eqnarray}
where $j$, $h$ and $C_3$ are functions of $t$, $\rho$, $z$ and $\phi$,
these equations become
\begin{eqnarray}
\label{phirhorho}
K_{\phi \rho \rho}:\, 0&=&  h_{,\rho} - h\alpha_{,\rho} - j \alpha_{,z} + \frac{1}{2} \rho^{-2}  e^{2\lambda} C_{1,\phi}\,,
\\
\label{phizz}
K_{\phi z z}:\, 0&=&  j_{,z} - h\alpha_{,\rho} - j \alpha_{,z} + \frac{1}{2} \rho^{-2}  e^{2\lambda} C_{1,\phi} \,,
\\
\label{phizrho}
K_{\phi z \rho}:\, 0&=&  h_{,z} + j_{,\rho} \,,
\\
\label{rhophiphi}
K_{\rho \phi \phi}:\, 0&=&  2 h_{,\phi} e^{2\nu} + \rho^2 C_{3,\rho} \,,
\\
\label{zphiphi}
K_{z\phi\phi}:\, 0&=&  2 j_{,\phi} e^{2\nu} + \rho^2 C_{3,z} \,,
\\
\label{phiphiphi}
K_{\phi \phi \phi}:\, 0&=&  h (\lambda_{,\rho} - \rho^{-1}) + j \lambda_{,z}
- \frac{1}{2} C_{3,\phi} - \frac{1}{2} \rho^{-2}  e^{2\lambda} C_{1,\phi} \,.
\end{eqnarray}
Combining Eq.\ (\ref{phizrho}) with the difference between 
Eqs.\ (\ref{phirhorho}) and (\ref{phizz}) we obtain the system
\begin{equation}
\label{nabla2}
j_{,\rho}+h_{,z}= 0 \,, \quad j_{,z} - h_{,\rho} =0 \,,
\end{equation}
which implies 
\begin{eqnarray}
j+ih=\mathcal{G}(\xi,\phi,t) \,,
\label{jhsolution}
\end{eqnarray}
where $\mathcal{G}(\xi,\phi, t)$ is an analytic function of $\xi$.
Substituting Eq.\ (\ref{jhsolution}) into the sum of Eqs.\ (\ref{phirhorho}) and (\ref{phizz}) gives
\begin{equation}
[V(\xi,{\bar \xi})\mathcal{G}(\xi,\phi,t)]_{,\xi} + ({\rm c.c.})
+C_{1,\phi}(\phi,t)V(\xi,{\bar \xi})\rho^{-2}e^{2\lambda(\xi,{\bar \xi})}=0 \,.
\end{equation}
Again, because $\lambda$ and $\nu$  are complicated non-analytic functions of $\rho$ and $z$, we conclude
that
\begin{equation}
\mathcal{G}(\xi,\phi,t)=0 \,, \quad C_{1,\phi}=0 \,, 
\label{calG0}
\end{equation}
which immediately gives
$j=h=0$ and thus $J=H=0 $. 
Accordingly,  Eqs.\ (\ref{rhophiphi}), (\ref{zphiphi}) and (\ref{phiphiphi}), imply $C_{3,\rho} = C_{3,z} = C_{3,\phi}  =0$, and thus that 
\begin{eqnarray}
\label{B}
B=C_1(t)\rho^2 e^{-2\lambda}+C_3(t)\rho^4e^{-4\lambda}.
\end{eqnarray}

We can now consider the remaining 10 Killing tensor equations.  
After changing variables to
\begin{eqnarray}
\label{change3}
A \equiv -C_1(t)e^{2\lambda}+C_2 e^{4\lambda} \,, \quad Q \equiv C_4 \rho^2 \,, \quad
F \equiv f e^{2\nu} \,, \quad G \equiv g e^{2\nu} \,,
\end{eqnarray}
where $C_2$, $C_4$, $f$ and $g$ are functions of $t$, $\rho$, $z$, and $\phi$, 
Eqs.\ (\ref{Krrt}) - (\ref{Ktpp}) take the form
\begin{eqnarray}
K_{\rho \rho t}:\, 0&=&  f_{,\rho}-f \alpha_{,\rho} - g \alpha_{,z} 
+\frac{1}{2}C_{1,t}e^{-2\lambda} \,, 
\label{rhorhot}
\\
K_{z z t}:\, 0&=&  g_{,z}- g \alpha_{,z} - f \alpha_{,\rho}  +\frac{1}{2}C_{1,t}e^{-2\lambda} \,,
\label{zzt}
\\
K_{\rho z t}:\, 0&=&  g_{,\rho}+f_{,z}  \,, 
\label{rhozt}
\\
K_{\rho t t}:\, 0&=&  2f_{,t} + e^{4 \lambda-2\nu} C_{2,\rho} \,,
\label{rhott}
\\
K_{ztt}:\, 0&=&  2g_{,t} + e^{4 \lambda-2\nu} C_{2,z} \,,
\label{ztt}
\\
K_{ttt}:\, 0&=&  f\lambda_{,\rho}+g\lambda_{,z} + \frac{1}{2} \left (
e^{-2 \lambda}C_{1,t} - C_{2,t} \right ) \,,
 \label{ttt}
\\
K_{\phi\rho t}:\, 0&=&   \rho^2 C_{4,\rho} + e^{2\nu} f_{,\phi} \,,
\label{phirhot}
\\
K_{\phi z t}:\, 0&=&  \rho^2 C_{4,z} + e^{2\nu} g_{,\phi} \,,
\label{phizt}
\\
K_{\phi t t}:\, 0&=&  2 \rho^2 C_{4,t} +e^{4 \lambda}C_{2,\phi}
\,,
\label{phitt}
\\
K_{t \phi \phi}:\, 0&=&  f \left (\lambda_{,\rho} -\rho^{-1} \right )
 +g\lambda_{,z}-  C_{4,\phi} 
 -\frac{1}{2}(C_{1,t} e^{-2\lambda}+C_{3,t}\rho^2 e^{-4\lambda}) \,.
\label{tphiphi}
\end{eqnarray}

Combining Eq.\ (\ref{rhozt}) with the difference between Eqs.\ (\ref{rhorhot}) and (\ref{zzt}) we obtain the system
\begin{equation}
\label{nabla2}
g_{,\rho}+f_{,z}= 0 \,, \quad  g_{,z}-f_{,\rho}  =0 \,,
\end{equation}
which implies 
\begin{eqnarray}
g+if=\mathcal{H}(\xi,\phi,t) \,,
\label{fgsolution}
\end{eqnarray}
where $\mathcal{H}(\xi, \phi,t)$ is an analytic function of $\xi$.
Substituting Eq.\ (\ref{fgsolution}) into the sum of Eqs.\ (\ref{rhorhot}) and (\ref{zzt}) gives
\begin{equation}
[V(\xi,{\bar \xi})\mathcal{H}(\xi,\phi,t)]_{,\xi} + ({\rm c.c.})
 +C_{1,t}(\phi,t)V(\xi,{\bar \xi})e^{-2\lambda(\rho,z)}=0 \,.
\end{equation}
As before, we conclude
that
\begin{equation}
\mathcal{H}(\xi,\phi,t)=0 \,, \qquad C_{1,t}=0 \, , 
\label{calH0}
\end{equation}
which immediately gives
$f=g=0$ and thus $F=G=0$.
Accordingly, Eqs.\ (\ref{rhott}) - (\ref{phizt}) imply
$C_{2,\rho} = C_{2,z} = C_{2,t} = 0$, and $C_{4,\rho} = C_{4,z} = 0$.  Equations (\ref{phitt}) and (\ref{tphiphi}) then yield
\begin{eqnarray}
2 \frac{\partial C_4(\phi, t)}{\partial t} + \rho^{-2} e^{4 \lambda} \frac{\partial C_2 (\phi)}{\partial \phi} &=& 0 \,,
\label{69}
\\
2 \frac{\partial C_4(\phi, t)}{\partial \phi} + \rho^{2} e^{-4 \lambda} \frac{\partial C_3 (t)}{\partial t} &=& 0 \,.
\label{70}
\end{eqnarray}
Eq.\ (\ref{69}) can be satisfied for all $\rho,\, z,\, \phi$ and $t$ if and only if $C_{4,t} = 0$ and $C_{2,\phi}=0$.  Similarly, Eq.\ (\ref{70}) can be satisfied for all $\rho,\, z,\, \phi$ and $t$ if and only if $C_{4,\phi} = 0$ and $C_{3,t}=0$.  
The final result is that $C_1$, $C_2$, $C_3$ and $C_4$ are arbitrary constants.

Combining these results, we find that the solution can be expressed in the form
\begin{equation}
\xi_{\alpha \beta}=  C_1 g_{\alpha \beta} + C_2 \xi^{(t)}_\alpha
\xi^{(t)}_\beta + C_3 \xi^{(\phi)}_\alpha
\xi^{(\phi)}_\beta - 2 C_4  \xi^{(t)}_{(\alpha}
\xi^{(\phi)}_{\beta )} \,,
\end{equation}
which is a linear combination of trivial Killing tensors.  We also note that this solution is obtained immediately using Maple.
We conclude that the Bach-Weyl spacetime does not admit a non-trivial second-rank Killing tensor, and thus, unlike its Newtonian limit, does not admit an analogous Carter-like constant of the motion.

\section{The post-Newtonian limit}

Here we show by explicit construction that, in the first post-Newtonian approximation of the Bach-Weyl metric, a Carter-like constant does not exist.  It suffices to note that the metric function $\nu$, being already quadratic in the masses, contributes to the motion only at 2PN order, and thus it can be set equal to zero in Eq.\ (\ref{metricds}).  The result is a metric in isotropic coordinates which matches the standard form, say of the parametrized post-Newtonian (PPN) framework~\cite{tegp}, with the GR values $\gamma=\beta=1$ of the PPN parameters.  Since the metric is static, 
the post-Newtonian equations of motion for a test body take the form
\begin{equation}
\frac{d^2 {\bm x}}{dt^2} = {\bm \nabla} U \left (1-4U+v^2 \right ) -4{\bm v} {\bm v} \cdot {\bm \nabla} U \,,
\end{equation}
where $U$ is the Newtonian gravitational potential given, in the Bach-Weyl case, by
$U = - \lambda =m_+/r_+ + m_-/r_- $.
Following the method used in the Newtonian Euler problem \cite{will09}, we calculate $dh^2/dt$, where ${\bm h} \equiv {\bm x} \times {\bm v}$: 
\begin{eqnarray}
\frac{1}{2} \frac{d h^2}{dt} &=& {\bm h} \cdot \left ({\bm x} \times \frac{d{\bm v}}{dt} \right )
\nonumber \\
&=&  {\bm h} \cdot  ({\bm x} \times {\bm \nabla} U ) \left (1-4U+v^2 \right ) -4h^2 \frac{d U}{dt} \,,
\label{carter1}
\end{eqnarray}
where we use the fact that, for a stationary potential, ${\bm v} \cdot {\bm \nabla} U = dU/dt$.  After some manipulation, it is straightforward to show that 
\begin{eqnarray}
{\bm h} \cdot  ({\bm x} \times {\bm \nabla} U )
&=&  \frac{d}{dt} (azU^*)  - a^2 v_z \nabla_z U 
\nonumber \\
&=& \frac{1}{2}  \frac{d}{dt} \left (2azU^* - a^2v_z^2 \right )
\nonumber \\
&&- a^2 v_z \left ( 4 v_z \frac{dU}{dt} + 4 U \nabla_z U - v^2 \nabla_z U \right ) \,,
\label{carter2}
\end{eqnarray}
where we used the PN equations of motion to go from $\nabla_z U$ in the first line to $dv_z/dt$,
and where $U^*  =m_+/r_+ - m_-/r_- $.
Considering just the Newtonian parts of Eqs.\ (\ref{carter1}) and (\ref{carter2}), we find the Newtonian Carter-like constant of the Euler problem, Eq.\ (\ref{sec1:carter}).

Now including the post-Newtonian terms in Eqs.\ (\ref{carter1}) and (\ref{carter2}), and using Newtonian equations of motion and the Newtonian Carter-like constant where necessary in those terms, we find
\begin{equation}
\frac{1}{2} \frac{d}{dt} \left [ (h^2 + a^2 v_z^2)(1+ 8 U)
-2azU^* (1+ v^2 +4U) \right ] = -6azU^* \frac{dU}{dt} \,.
\end{equation}
The term on the right-hand side cannot be expressed as a total time derivative of any function of $\bm x$ or $\bm v$, and so a Carter-like constant does not exist at post-Newtonian order.

\section{Discussion}

We have shown that a Carter-like constant of the motion analogous to the one that exists in the Newtonian Euler problem does not exist in the relativistic analogue given by the Bach-Weyl solution, both by showing the non-existence of a non-trivial second-rank Killing tensor and by showing the absence of such a constant at post-Newtonian order.  However, this does not completely rule out the integrability of motion in the Bach-Weyl spacetime, because there could in principle exist a higher-order Killing tensor, or equivalently an additional constant of motion more complicated than the Carter-like constant we have considered.  

It is also worth pointing out that, because our solution did not depend crucially on the specific functional form of the function $V$, it applies to other solutions in the Bach-Weyl class, such as the two black hole solution (corresponding to a potential $\lambda$ generated by two collinear rods on the $z$-axis) discussed in \cite{coelho}.  In that solution, even the Newtonian limit fails to have a Carter-like constant.

\ack
This work was supported in part by the National Science Foundation, Grant Nos.\ PHY 06--52448 and 09--65133, the National Aeronautics and Space Administration, Grant No.\
NNG-06GI60G, and the Centre National de la Recherche Scientifique, Programme Internationale de la Coop\'eration Scientifique (CNRS-PICS), Grant No. 4396.  We are grateful to Sterl Phinney, Donald Lynden-Bell, Jean-Philippe Uzan, Stanley Sheppard and Diarmuid Math\'una for useful information about the Euler problem, and to Jeandrew Brink for a useful discussion about Killing tensors.  We are especially grateful to David Garfinkle for important suggestions about solving the Killing tensor equations.

\section*{Appendix:  On the vanishing of the analytic function ${\cal F}$}

In this Appendix we present an argument in support of the claim (\ref{calF0}) that the analytic function $\cal F$ must vanish.  
Eq.\ (\ref{Fequation}) can also be written as a real differential equation in terms of $Y$ and $X$, in the form 
\begin{equation}
2VY_{,\rho z} + 3(Y_{,\rho} V_{,z} +Y_{,z} V_{,\rho} ) + 2Y V_{,\rho z} + X (V_{,\rho\rho} - V_{,zz}) =0 \,.
\label{maineq}
\end{equation}
We perform a Taylor expansion of the functions $X$, $Y$, and $V$ about an arbitrary point $(\rho_0, z_0)$, up to an order $M$, corresponding to the powers $\Delta \rho^q \Delta z^{M-q}$, where $q = 0 \dots M$,  $\Delta \rho = \rho - \rho_0$, and $\Delta z = z-z_0$.  Now because the function ${\cal F} = Y + iX$ is analytic, it can be expanded in the form
\begin{equation}
{\cal F} = \sum_{n=0}^\infty (a_n + ib_n) (\Delta z + i\Delta \rho)^n /n! \,,
\end{equation}
where $a_n$ and $b_n$ are real coefficients to be determined.  Up to a given order $M$, we need $2(M+1)$ coefficients to specify the expansion of $\cal F$ and hence of $X$ and $Y$.  We also expand the function $V$ about $(\rho_0, z_0)$, and then substitute these expansions into the differential equation (\ref{maineq}).  For each order $M$ in the expanded differential equation, there is a homogeneous algebraic equation for the coefficient of $\Delta \rho^q \Delta z^{M-q}$, thus there are $M+1$ separate equations.  Consequently the total number of algebraic equations from $M=0$ up to and including order $M$ is 
\begin{equation}
N = \frac{1}{2} (M+1)(M+2) \,.
\end{equation}
However, because Eq.\ (\ref{maineq}) is a second order differential equation, at a given order $M$, coefficients up to order $M+2$ in $X$ and $Y$ will appear.  Therefore there will be $P= 2(M+3)$ expansion coefficients of $\cal F$ to be fixed at $(\rho_0,\, z_0)$ via this system of linear algebraic equations.

For small $M$, $N < P$, and so the linear equations have solutions for any function $V$.  But $N$ grows quadratically with $M$, while $P$ grows linearly, and eventually there will be more equations than there are coefficients of $X$ and $Y$.  Then, solutions for the coefficients of $\cal F$ will exist only if the coefficients of the expansion of $V$ satisfy conditions.   The cross-over occurs when $M =4$; in this case, the number of equations is 15, while the number of coefficients is only 14.   Setting one of the equations aside for a moment,
for the $14 \times 14$ system of linear equations, there will be a non-zero solution for the coefficients of $\cal F$ if and only if the determinant of the matrix vanishes.  The matrix depends only on the value of $V$ and its derivatives up to order $M+2=6$ at 
$(\rho_0, z_0)$.  For a generic $V$, the determinant will not vanish, and so the only solution is for all $a_n$ and $b_n$ to vanish for $n \le M+2$.  But since the expansion of the differential equation gives the $a_n$ and $b_n$ for $n > M+2$ in terms of the values for $n \le M+2$, vanishing of the latter implies vanishing of all coefficients.  Since $(\rho_0, z_0)$ was arbitrary, this implies that ${\cal F} = 0$ everywhere.  
Even if the expansion coefficients of $V$ manage to satisfy the conditions at order $M=4$, they must continue to do so order by order.  As a result, there will be a 
non-zero solution for $\cal F$ if and only if $V$ is extraordinarily fine tuned.   The same argument can be applied to support the vanishing of the analytic functions $\cal G$ and $\cal H$ in Eqs.\ (\ref{calG0}) and (\ref{calH0}).

How finely tuned must $V$ be?  
In spherical symmetry, where  $V = V(\rho^2+z^2)= V(\xi \bar{\xi})$
and $\nu = 0$ it is simple to show that ${\cal F} = \xi^2$ is a solution to Eq.\ (\ref{Fequation}).  In this case the solution merely generates additional trivial Killing tensors coming from products involving the two additional rotational Killing vectors that exist in spherical symmetry (giving a total of 11 trivial Killing tensors).  The Bach-Weyl solution completely lacks this additional symmetry.

\end{document}